\documentclass[onecolumn, floatfix, preprintnumbers, eqsecnum, letterpaper, superscriptaddress, nofootinbib]{revtex4}
\usepackage{graphicx}
\usepackage{microtype}
\usepackage{amsmath}
\usepackage{amssymb}
\usepackage{subfigure}
\usepackage{hyperref}
\usepackage{url}
\usepackage{xcolor}
\usepackage{color}
\usepackage{mathrsfs}
\usepackage{calrsfs}
\usepackage{amsfonts}
\usepackage{eufrak}
\usepackage{tabularx}
\usepackage{eucal}
\usepackage{latexsym}
\usepackage{ragged2e}
\usepackage{epsfig}
\usepackage{textcomp}
\usepackage{inputenc}

\usepackage{caption}
\usepackage{subfigure}

\usepackage{marvosym}
\usepackage{ifsym}

\usepackage{lipsum}

\DeclareCaptionJustification{justified}{\leftskip=0pt \rightskip=0pt \parfillskip=0pt plus 1fil}
\definecolor{vividviolet}{rgb}{0.62, 0.0, 1.0}
\definecolor{amaranth}{rgb}{0.9, 0.17, 0.31}
\definecolor{palatinateblue}{rgb}{0.15, 0.23, 0.89}
\definecolor{brightpink}{rgb}{1.0, 0.0, 0.5}
\definecolor{cornflowerblue}{rgb}{0.39, 0.58, 0.93}
\definecolor{deepcarminepink}{rgb}{0.94, 0.19, 0.22}
\definecolor{radicalred}{rgb}{1.0, 0.21, 0.37}

\hypersetup{ linktoc=all,
	colorlinks, linkcolor={palatinateblue},
	citecolor={brightpink}, urlcolor={amaranth}
}

\graphicspath{{Images/}}

\graphicspath{{Images/}}

\def\@fnsymbol#1{\ensuremath{\ifcase#1\or \ddagger \or  $\textleaf$  \or \dagger
		\else\@ctrerr\fi}}%

\makeatother

\def\sideremark#1{\ifvmode\leavevmode\fi\vadjust{\vbox to0pt{\vss
			\hbox to 0pt{\hskip\hsize\hskip1em
				\vbox{\hsize1.3cm\tiny\raggedright\pretolerance10000
					\noindent #1\hfill}\hss}\vbox to8pt{\vfil}\vss}}}%
\def\beq{\begin{equation}}
	\def\eeq{\end{equation}}

\begin{document}
\title{Gravitational lensing by a charged spherically symmetric \\ black hole immersed in thin dark matter}

\author{Xiao-Jun Gao}\email{xjgao2020@nuaa.edu.cn}
\address{College of Physics, Nanjing University of Aeronautics and Astronautics, Nanjing 211106, China}

\author{Xiao-kun Yan}\email{yanxiaokun2008035@nuaa.edu.cn}
\address{College of Physics, Nanjing University of Aeronautics and Astronautics, Nanjing 211106, China}

\author{Yihao Yin}\email{yinyihao@nuaa.edu.cn}
\address{College of Physics, Nanjing University of Aeronautics and Astronautics, Nanjing 211106, China}
\address{Key Laboratory of Aerospace Information Materials and Physics (NUAA), MIIT, Nanjing 211106, China}

\author{Ya-Peng Hu}\email{huyp@nuaa.edu.cn}
\address{College of Physics, Nanjing University of Aeronautics and Astronautics, Nanjing 211106, China}
\address{Key Laboratory of Aerospace Information Materials and Physics (NUAA), MIIT, Nanjing 211106, China}

\begin{abstract}

We investigate the gravitational lensing effect around a spherically symmetric black hole, whose metric is obtained from the Einstein field equation with electric charge and perfect-fluid dark matter contributing to its energy-momentum tensor. We do the calculation analytically in the weak field limit and we assume that both the charge and the dark matter are much less abundant (only give rise to the next-leading-order contribution) in comparison to the black hole mass. In particular, we derive the light deflection angle and the size of the Einstein ring, where approximations up to the next-leading order are done with extra care, especially for the logarithmic term from perfect-fluid dark matter. We expect our results will be useful in the future to relate the theoretical model of perfect fluid dark matter with observations of celestial bodies immersed in thin dark matter.


\end{abstract}

\maketitle



\section{Introduction}
Although dark matter has been difficult to directly detect due to its lack of electromagnetic interaction, abundant observational evidences indicate that typical galaxies are filled with a great deal of it (see e.g.\ \cite{Rubin:1980zd,Persic:1995ru,Bertone:2016nfn}). Thus when we study celestial bodies, say black holes, in realistic astrophysical settings, we should think them as objects immersed in dark matter for better accuracy.
In this scenario to obtain the metric, one can look for black hole solutions to the Einstein field equations where the energy-momentum tensor is contributed by dark matter.
The simplest solutions are the spherically symmetric ones where only perfect fluid dark matter (PFDM) is involved. For instance, this kind of solutions have been obtained in \cite{Kiselev:2002dx} and \cite{Li:2012zx} where the dark matter are assumed to be the quintessence scalar field and weakly interacting massive particles, respectively, and in both cases the solutions are characterized by a logarithmic term in the metric function. These solutions have further been extended to the Kerr situation~\cite{Toshmatov:2015npp,Xu:2017bpz}. Spacetime properties of the above-mentioned solutions have been extensively investigated in~\cite{Haroon:2018ryd,Hou:2018avu,Xu:2016ylr,Ma:2020dhv,Rayimbaev:2021kjs,Shaymatov:2020wtj,Atamurotov:2021hoq,Atamurotov:2021hck,Das:2020yxw,Shaymatov:2020bso,Das:2021otl,Shaymatov:2021nff}.
Gravitational lensing (GL) has been a well-known phenomenon that can be used to relate theoretical analysis of black hole spacetime with its astronomical observations, and thus it would be interesting to investigate GL effects of a black hole immersed in PFDM.

In the literature, at least in the weak field approximation, there have been two methods to calculate the light deflection angle (mainly in the context without dark matter). The traditional method is to obtain the path of a light ray by solving the geodesic equations~\cite{Sereno:2003nd,Keeton:2005jd,Hu:2013eya,Gao:2019pir,Gao:2021lmo,Fu:2021fxn,Virbhadra:2022iiy,Virbhadra:2022ybp,Liu:2022lfb}.
The newer method, which explicitly exhibits the relation between the deflection angle and the Gaussian curvature, was proposed by Gibbons and Werner based on the Gauss-Bonnet theorem using the optical metrics~\cite{Gibbons:2008rj,Werner:2012rc},  and for recent applications see ~\cite{Jusufi:2017lsl,Jusufi:2017hed,Jusufi:2017uhh,Jusufi:2017mav,Jusufi:2018jof,Ovgun:2018xys,Javed:2019qyg,Fu:2021akc,Gao:2022cds}. In the literature, the calculation usually involves the approximation that both the light source and the receiver are infinitely far from the black hole. It was not until recently that finitely distant endpoints of the light path have been considered~\cite{Ishihara:2016vdc,Ono:2017pie,Takizawa:2020egm,Li:2020wvn}.

Using the Gauss-Bonnet theorem, a few pioneering works have been done on the calculation of the light deflection angle around a black hole immersed in PFDM. The first one was \cite{Haroon:2018ryd}, where the light deflection angle was calculated for a rotating black hole in PFDM, and later in \cite{Atamurotov:2021hck} the charge of the black hole was also taken into account and the angular radius of the Einstein ring was further calculated. In these works, the weak field approximation has been adopted, i.e.\ the ratio between the mass parameter $M$ and the impact parameter $b$ has been assumed to be much smaller than $1$, and the calculation has been merely up to the first order of $M/b$. Furthermore, the calculation therein has also exploited series expansions with respect to several parameters, e.g.\ the PFDM parameter $\lambda$ and the charge parameter\footnote{To clarify our notation, $q$ is defined for simplicity as a parameter proportional to the square of the electric charge.} $q$, and only their leading order has been retained as an approximation.

It is important to notice that, such approximation has an underlining presumption: the leading-order contributions to the result from $\lambda$ and $q$ are of the same order of magnitude as that from $M$, which in more precise words means: using $b$ to nondimensionalize the parameters, $\mathcal{O}(M/b) \sim \mathcal{O}(\lambda/b) \sim \mathcal{O}(q/b^2)$ has been presumed in the above mentioned works (not explicitly stated though). We think such presumption is defective in the following sense. First, although galaxies are usually dominated by dark matter, there is no reason to believe that a single celestial object like a regular star or black hole in a typical astronomical environment is that heavily surrounded by dark matter. Second, as we have learned from Reissner-Nordstr\"om black hole, the charge of a black hole has an upper limit set by its mass, otherwise it becomes a naked singularity. Therefore, in many realistic astronomical environments it is more likely that $M/b \gg \lambda/b$ (i.e.\ what we mean by ``thin dark matter'' in the title) and that $M/b \gg q/b^2$, and then the calculation makes little sense if one drops the next leading order of $M$, since it may be as large as the leading order of $\lambda$ and $q$.

In this paper we will investigate GL around a charged spherically symmetric black hole immersed in thin dark matter, in particular with the presumption $\mathcal{O}(M^2/b^2) \sim \mathcal{O}(\lambda/b) \sim \mathcal{O}(q/b^2)\ll \mathcal{O}(M/b)\ll 1$. We will analytically derive the light deflection angle, with the calculation pushed to the next leading order i.e.\ $\mathcal{O}(M^2/b^2)$.
In addition, we will further investigate the size of the Einstein ring, which we think may be useful in the future to relate the result of this paper with astronomical observations. Note that approximations in this paper are done in a much more careful manner in comparison to the literature.
For instance, in the literature e.g.\ \cite{Atamurotov:2021hck} the light source and receiver were usually treated as infinitely far from the lensing object, but in this paper, we will not do this approximation until we finish deriving the deflection angle, and specify how far is necessary for infinity to be a good approximation.\footnote{Note that in the literature, Ref.~\cite{Haroon:2018ryd} was an exception that the source and receiver were treated as finitely distant, but the calculation there did not involve $\mathcal{O}(M^2/b^2)$ in comparison to this paper.} For another example, the appearance of $\ln(b/\lambda)$-terms in the formulas are usually the biggest difficulty in doing analytical calculations, and in \cite{Atamurotov:2021hck} to derive the size of the Einstein ring, the problem was evaded in an unnatural way that $b/\lambda$ is replaced with $10^n$, where $n$ is a new constant introduced by hand. In Sec.~\ref{section4} of this paper, we will instead resolve such difficulty by a different approximation that is based on a more solid analysis of the orders of small quantities.

Our paper is organized as follows: In Sec.~\ref{section2}, we first do a brief review on the relation between the light deflection angle and the Gauss-Bonnet theorem for the light source and receiver at finite distance. In Sec.~\ref{section3}, using the Gauss-Bonnet theorem, we calculate the weak deflection angle around a charged spherically symmetric black hole immersed in PFDM. In Sec.~\ref{section4}, we derive the the Einstein angular radius in this scenario. Finally Sec.~\ref{section5} is for conclusion and discussion. Throughout this paper we assume asymptotic flatness of spacetime and we use the geometric units with $G=c=1$ unless otherwise specified.

\section{Review: the relation between Gauss-Bonnet theorem and deflection angle in finite distance}
\label{section2}
In a spherically symmetric spacetime background, e.g.\ around a non-rotating (charged) black hole, it is obvious that the light travels within an equatorial plane, and thus it is interesting to write down an equation that directly relates the deflection angle of the light with the curvature of the equatorial plane. The Gauss-Bonnet theorem offers such a direct relation, which we will briefly review in this section.

The line element of a static four-dimensional spherically symmetric spacetime can be written as
\begin{align}
 ds^2=-A(r)dt^2+B(r)dr^2+r^2d\Omega^2, \label{4Dspacetime}
\end{align}
where $d\Omega^2\equiv d\theta^2+\sin^2\theta d\phi^2$.
For the trajectory of the light, we impose the null condition $ds^2=0$, which leads to
\begin{align}
dt^2=\gamma_{ij}{dx^{i}dx^{j}}=\dfrac{B(r)}{A(r)}dr^2+\dfrac{r^2}{A(r)} d\Omega^2, \label{opticalmetric}
\end{align}
where $\gamma_{ij}$ is called the optical metric, and $i$ and $j$ run from $1$ to $3$. The optical metric can define a three-dimensional Riemannian space, whose coordinates is used to describe light rays.

Without losing generality, we choose the equatorial plane $\theta=\pi/2$ to investigate the path of the light. One can use the conserved energy ($E$) and angular momentum ($L$) of light in a static spherically symmetric spacetime to define the impact parameter~\cite{Ishihara:2016vdc}
\begin{align}
b\equiv\dfrac{L}{E}=\dfrac{r^2}{A(r)}\dfrac{d\phi}{dt}.\label{impactparameter}
\end{align}
From the above equations one can further derive \cite{Hu:2013eya,Ishihara:2016vdc}
\begin{align}
\left(\dfrac{du}{d\phi}\right)^2=\dfrac{1}{b^2 A(u) B(u)}-\dfrac{u^2}{B(u)}\equiv F(u),\label{trajectory2}
\end{align}
where $u \equiv 1/r$, and for convenience later we will use the notations $u_S$ and $u_R$ for the value of $u$ at the source and the receiver, and $u_0$ is that of the closest point the photon's trajectory to the black hole.
In the context of a finitely distant source and receiver, the light deflection angle is given by  ~\cite{Ishihara:2016vdc}
\begin{align}
\hat{\alpha}\equiv\Psi_R-\Psi_S+\phi_{RS},\label{definitionangle}
\end{align}
where $\phi_{RS}$ is the coordinate angle between the two radial directions of the receiver and the source:
\begin{align}
\phi_{RS}=\int_{u_S}^{u_0}\dfrac{du}{\sqrt{F(u)}}+\int_{u_0}^{u_R}\dfrac{du}{\sqrt{F(u)}},\label{phiRS}
\end{align}
and $\Psi_S$ ($\Psi_R$) is the angle between the light's trajectory and the radial direction of the source (receiver).
Furthermore, following the notation of~\cite{Ishihara:2016vdc}, for each point of the light's trajectory, the angle between its tangent direction and the radial coordinate is denoted by $\Psi$, which satisfies
\begin{align}
\sin\Psi=\dfrac{b\sqrt{A(r)}}{r}.\label{sinPsi}
\end{align}

\begin{figure}[htbp]
\begin{minipage}[t]{0.48\linewidth}
\centering
\includegraphics[width=3.2in]{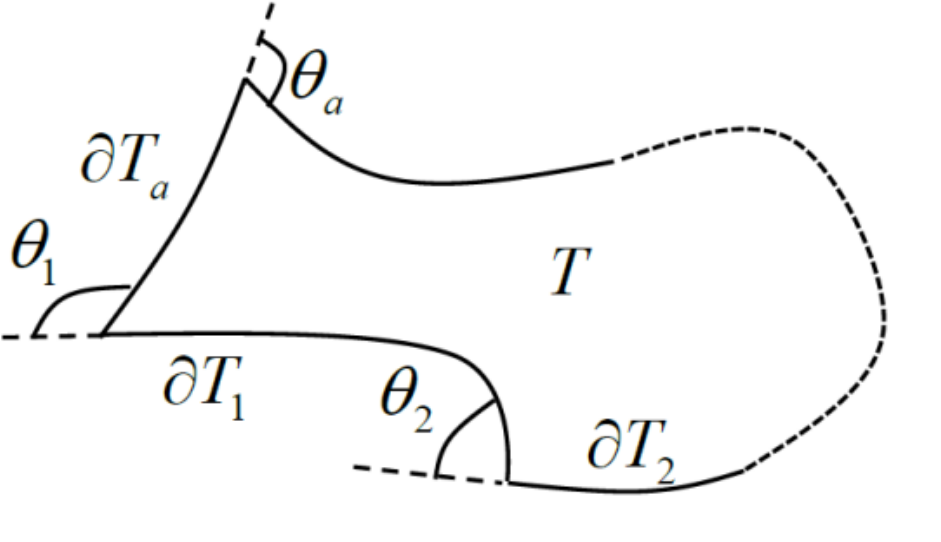}
\caption{Schematic figure for the Gauss-Bonnet theorem (adopted from Ref.~\cite{Takizawa:2020egm}).}
\label{fig:side:a}
\end{minipage}
\hspace{0.1cm}
\begin{minipage}[t]{0.48\linewidth}
\centering
\includegraphics[width=3.2in]{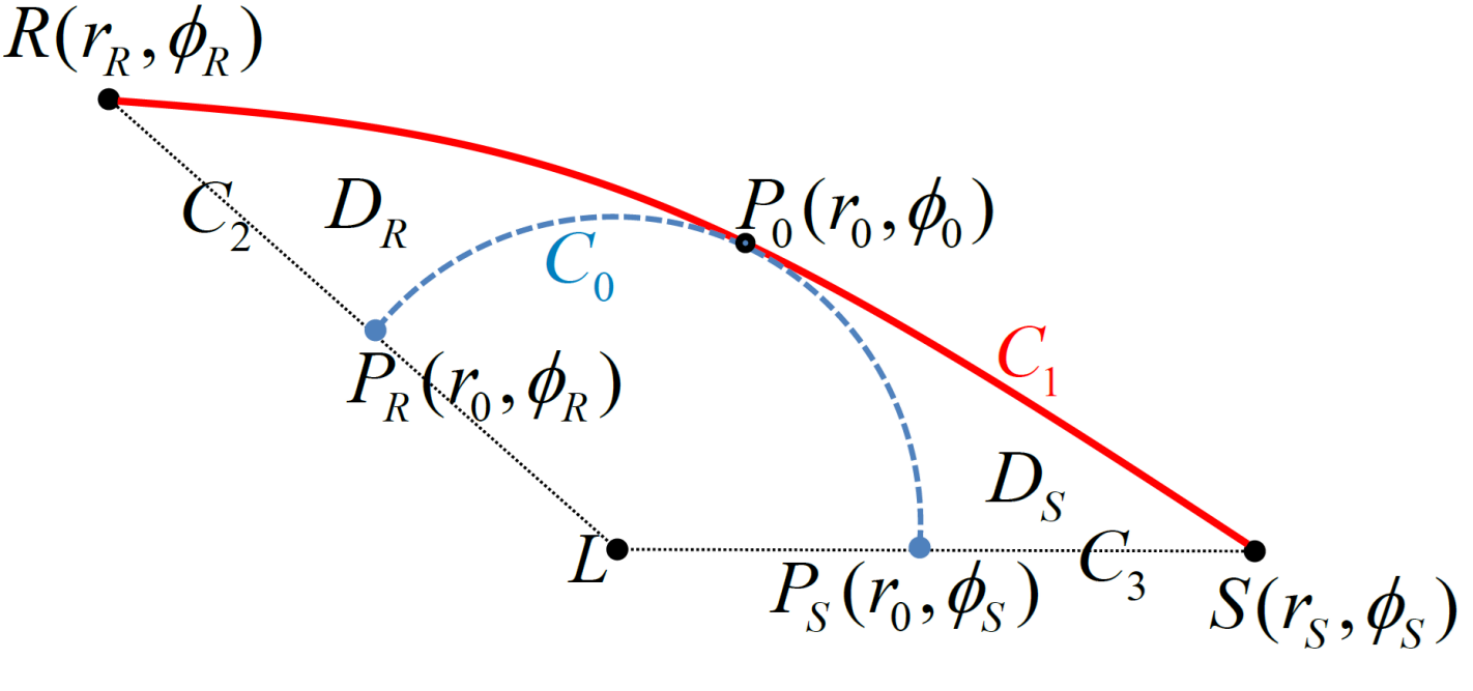}\\
\caption{Schematic figure for the regions $D_R$ and $D_S$ (adopted from Ref.~\cite{Takizawa:2020egm}).}
\label{fig:side:b}
\end{minipage}
\end{figure}

The Gauss-Bonnet theorem reveals the relation between the intrinsic differential geometry and topology of the surface. The domain $T$ be a compact oriented nonsingular two-dimensional Riemannian surface with Euler characteristic $\chi(T)$ and Gaussian curvature $K$. Its boundary $\partial T$ is a piecewise smooth curve. The Gauss-Bonnet theorem can be expressed as~\cite{GBMath,Landau}
\begin{align}
\iint_{T}KdS+\oint_{\partial T}k_{g}dl+\sum\theta_a=2\pi\chi(T),\label{GaussBonnettheorem}
\end{align}
where $dS$ is the area element of the surface
\begin{align}
dS=\sqrt{\gamma_{rr}\gamma_{\phi\phi}}drd\phi \  ; \label{areaelement}
\end{align}
$dl$ denotes the line element along the boundary; $\theta_a$ stands for the external angle at $a$th vertex (see Fig.\ref{fig:side:a}); $K$ is the Gaussian curvature of
the optical space written as~\cite{Li:2019vhp}
\begin{align}
K=\dfrac{R_{r\phi r\phi}}{\det(\gamma_{ij})}=&-\dfrac{1}{\sqrt{\gamma_{rr}\gamma_{\phi\phi}}}\left[\dfrac{\partial}{\partial r}\left(\dfrac{1}{\sqrt{\gamma_{rr}}}\dfrac{\partial\sqrt{\gamma_{\phi\phi}}}{\partial r}\right)+\dfrac{\partial}{\partial \phi}\left(\dfrac{1}{\sqrt{\gamma_{\phi\phi}}}\dfrac{\partial\sqrt{\gamma_{rr}}}{\partial \phi}\right)\right] \ ; \label{Gaussiancurvature}
\end{align}
and $k_g$ is the geodesic curvature of a smooth curve $C$ defined by $r(\phi)$. From Fig.\ref{fig:side:b}, $C=r(\phi)=C_0=$ constant, $k_g$ is expressed~\cite{Li:2019vhp}:
\begin{align}
k_g(C_0)=|\nabla_{\dot{C_0}}\dot{C_0}|=\Gamma^r_{\phi\phi}(\dot{C}_0^\phi)^2,\label{geodesiccurvature}
\end{align}
where $\dot{C_0}$ denotes the tangent vector along the smooth curve $C_0$, and $\Gamma^r_{\phi\phi}$ is the Christoffel symbol. One easily find that the $\dot{C}_0^\phi$ can be calculated via the unit speed condition, i.e.\ $\gamma_{\phi\phi}\dot{C}^\phi_0\dot{C}^\phi_0=1$. But if the curve $C$ is a spatial geodesic leading to $k_g(C)=0$.

Takizawa $et$ $al.$ ~\cite{Takizawa:2020egm} considered a specific region shown in Fig.\ref{fig:side:b}. $L$ is the center of the black hole, i.e.\ the origin of the coordinates. The boundary of the region consists of following: $C_1$ is a spatial geodesic from the source $S$ to the receiver $R$; $C_0$ is the circular arc centered around $L$ and tangent to $C_1$; $C_2$ and $C_3$ are the radial lines that start from $P_R$ and $P_S$, respectively, on $C_0$ and end at $R$ and $S$ on $C_1$. Thus the region is divided into two domains $D_R$ and $D_S$, then by using the Gauss-Bonnet theorem (\ref{GaussBonnettheorem}) one obtains
\begin{align}
\iint_{D_R}KdS+\int^{P_0}_{P_R}k_{g}(C_{0})dl-\Psi_R+\dfrac{\pi}{2}=0\label{RGaussBonnettheorem}
\end{align}
and
\begin{align}
\iint_{D_S}KdS+\int^{P_S}_{P_0}k_{g}(C_{0})dl+\Psi_S-\dfrac{\pi}{2}=0,\label{SGaussBonnettheorem}
\end{align}
where the facts that $k_g(C_1)=k_g(C_2)=k_g(C_3)=0$ and that the inner angle at the tangent point is zero have been used. By using (\ref{RGaussBonnettheorem}) and (\ref{SGaussBonnettheorem}) to eliminate the $\Psi_R$ and $\Psi_S$ in  (\ref{definitionangle}), the expression of the deflection angle can be rewritten in a way that explicitly depends on the Gaussian curvature
\begin{align}
\hat{\alpha}=\iint_{D_R+D_S}KdS+\int^{P_S}_{P_R}k_{g}(C_0)dl+\phi_{RS}.\label{finallyangle}
\end{align}
In the next section, we will apply the (\ref{finallyangle}) to the calculation of the weak deflection angle of light for the charged black hole immersed in PFDM.

\section{Light deflection angle around a charged spherically black hole immersed in PFDM}
\label{section3}
In this section, we first briefly review the static spherically symmetric charged black hole solution to the Einstein field equation with PFDM, and then we calculate the weak deflection angle for a pair of finitely distant light source and receiver using the method of \cite{Takizawa:2020egm} based on Gauss-Bonnet theorem.

\subsection{Static spherically symmetric black hole metric with charge and PFDM}
The action of Einstein-Maxwell gravity in the presence of dark matter is written as~\cite{Xu:2016ylr,Atamurotov:2021hck,Das:2020yxw}:
\begin{align}
S=\int d^4x\sqrt{-g}\left[\dfrac{\bar{R}}{16 \pi}+\dfrac{1}{4}F^{\mu\nu}F_{\mu\nu}+ \mathcal{L}_{\text{DM}}\right],\label{darkmatteraction}
\end{align}
where $\bar{R}$ denotes the Ricci scalar; $F_{\mu\nu}=\partial_\mu A_\nu-\partial_\nu A_\mu$ is the electromagnetic field;
and $\mathcal{L}_{\text{DM}}$ is the dark-matter Lagrangian. By variation of this action about the metric $g_{\mu\nu}$, the Einstein field equation is obtained as follows
\begin{align}
\bar{R}_{\mu\nu}-\dfrac{1}{2}g_{\mu\nu}\bar{R}&=8\pi \bar{T}_{\mu\nu},\label{DMEinsteinequation}
\end{align}
where the energy momentum tensor contains two parts:
\begin{equation}
\bar{T}_{\mu\nu}= T_{\mu\nu}^{M}-T_{\mu\nu}^{DM} \ .
\end{equation}
Here $T^{DM}_{\mu\nu}$ is the energy-momentum tensor of the dark matter, which as a perfect fluid can be written as \cite{Das:2020yxw}
\begin{align}
(T^{\mu}~_{\nu})^{DM}=\ {\rm diag}\left(-\rho,P_r,P_\theta,P_\phi\right); ~\rho=-P_r; ~P_\theta=P_\phi,\label{PFDMtensor}
\end{align}
where $\rho$ and $P_i$ correspond to the density and pressure; and $T_{\mu\nu}^{M}$ is the energy-momentum tensor of the electromagnetic field:
\begin{equation}
T^M_{\mu\nu}=\dfrac{1}{4\pi}\left(F_{\mu\sigma} F_\nu~^\sigma-\dfrac{1}{4}g_{\mu\nu}F_{\sigma\tau}F^{\sigma\tau}\right).
\label{electromagnetictensor}
\end{equation}

For static spherically symmetric solutions to (\ref{DMEinsteinequation}), one can write down the following ansatz:
\begin{align}
ds^2=-f(r)dt^2+\dfrac{1}{f(r)}dr^2+r^2(d\theta^2+\sin^2\theta d\phi^2)\label{ansatzmetric},
\end{align}
while adopting
\begin{equation}
   A_\mu=(-\frac{Q}{r},0,0,0) \label{Amu}
\end{equation}
as the simplest electrostatic solution to the Maxwell equations, where $Q$ is the electric charge. From (\ref{Amu}) one can derive that
\begin{align}
F_{rt}=-F_{tr}=\dfrac{Q}{r^2},\label{Frtvalue}
\end{align}
while other components of $F_{\mu\nu}$ all vanish.
Then by using (\ref{ansatzmetric}) and (\ref{Frtvalue}), one derives from (\ref{electromagnetictensor}) that
\begin{align}
T_{\mu\nu}^M=\dfrac{Q^2}{8\pi r^4}\ {\rm diag}\left(f(r),-\dfrac{1}{f(r)},r^2,r^2\sin^2\theta\right).\label{chargetensor}
\end{align}
Using the ansatz (\ref{ansatzmetric}) together with (\ref{PFDMtensor}) and (\ref{chargetensor}), one then solve the Einstein field equation, which gives
\begin{align}
f(r)=1-\dfrac{2M}{r}+\dfrac{q}{r^2}+\dfrac{\lambda}{r}\ln\left(\dfrac{r}{\lambda}\right),\label{RNdSPFDMmetric}
\end{align}
where for simplicity we denote $q=Q^2$; $M$ is the mass of the black hole; and $\lambda$ parametrizes the dark matter and is related to the perfect fluid density and pressure by
\begin{align}
\rho=-P_r=\dfrac{\lambda}{8 \pi r^3},~P_\theta=P_\phi=\dfrac{\lambda}{16 \pi r^3}\label{energydensity}
\end{align}
and thus $\lambda>0$ due to the weak energy condition $\rho> 0$. For details of the above derivation, see Ref.~\cite{Das:2020yxw}.

\subsection{Calculation of deflection angle with the Gauss-Bonnet theorem}
Fitting (\ref{ansatzmetric}) into the form of (\ref{4Dspacetime}), we obtain
\begin{align}
A(r)=B(r)^{-1}=f(r).\label{4DEGBmetric-new}
\end{align}
Then the optical metric on the equatorial plane ($\theta=\pi/2$) is given by
\begin{align}
\gamma_{rr}=\left[1-\dfrac{2M}{r}+\dfrac{q}{r^2}+\dfrac{\lambda}{r}\ln\left(\dfrac{r}{\lambda}\right)\right]^{-2},~~
\gamma_{\phi\phi}=r^2\left[1-\dfrac{2M}{r}+\dfrac{q}{r^2}+\dfrac{\lambda}{r}\ln\left(\dfrac{r}{\lambda}\right)\right]^{-1}.
\end{align}
Thus the Gaussian curvature (\ref{Gaussiancurvature}) can be written as
\begin{align}
K=\dfrac{R_{r\phi r\phi}}{det(\gamma_{ij})}=&-\dfrac{1}{\sqrt{\gamma_{rr}\gamma_{\phi\phi}}}\left[\dfrac{\partial}{\partial r}\left(\dfrac{1}{\sqrt{\gamma_{rr}}}\dfrac{\partial\sqrt{\gamma_{\phi\phi}}}{\partial r}\right)+\dfrac{\partial}{\partial \phi}\left(\dfrac{1}{\sqrt{\gamma_{\phi\phi}}}\dfrac{\partial\sqrt{\gamma_{rr}}}{\partial \phi}\right)\right]\notag\\
=&\frac{u^3}{4} \{ [12 M^2 u-8 M (3 q u^2-\lambda  u+1)+(2 q u-\lambda ) (u (\lambda +4 q u)+6)]\notag\\
&+ \lambda \ln (\lambda  u) [4 [u (\lambda +3 M-3 q u)-1]+3 \lambda  u \ln (\lambda  u)]\}\label{KRNdSPFDM}
\end{align}
and the area element (\ref{areaelement}) as
\begin{align}
dS=\sqrt{det(\gamma_{ij})}drd\phi=-u^{-3} \sqrt{\frac{1}{\left[1-2 M u+q u^2-\lambda  u \ln (\lambda  u)\right]^3}}du d\phi,\label{elementRNdSPFDM}
\end{align}
where $r$ has been replaced with $1/u$. By substituting (\ref{4DEGBmetric-new}) into the equation of trajectory (\ref{trajectory2}), it can be easily obtained that
\begin{align}
\left(\dfrac{du}{d\phi}\right)^2=\dfrac{1}{b^2}-u^2+2Mu^3+\lambda u^3\ln\left(\lambda u\right)-qu^4.\label{RNdSPFDMtrajectory}
\end{align}

Due to the complexity of the (\ref{RNdSPFDMtrajectory}), its exact solution can be difficult to obtain, and therefore in the following we proceed with the approximation in the weak field limit, i.e.\ writing formulas only in leading orders of the dimensionless quantity $M/b\ll 1$.
Recall that for the existence of the event horizon of the Reissner-Nordstr\"om black hole~\cite{Eiroa:2002mk} the charge should not be too large comparing to $M$, and thus it is reasonable to assume $q/b^2\ll M/b$. Furthermore, in the context of this paper, we assume that the thin dark matter affects spacetime much less than the black hole mass does, i.e.\ $\lambda/b\ll M/b$. Taking all these into account and to simplify our discussion, we denote $\varepsilon$ as a dimensionless small quantity such that $\mathcal{O}(\varepsilon) \sim \mathcal{O}(M/b)$, and in this paper we further assume that $\mathcal{O}(\lambda/b)\sim \mathcal{O}(q/b^2)\sim \mathcal{O}(M^2/b^2)\sim \mathcal{O}(\varepsilon^2)\ll\mathcal{O}(M/b)\ll 1$. In the following , our calculations will be precise up to the second-order $\mathcal{O}(\varepsilon^2)$.

From (\ref{RNdSPFDMtrajectory}), we can perturbatively obtain
\begin{align}
u(\phi)=&\dfrac{1}{b}\left\{\sin\phi+\frac{M}{b}(1+\cos^2\phi)+\frac{M^2[37 \sin\phi-3 \sin3 \phi+30 (\pi -2 \phi ) \cos \phi]}{16 b^2}-\dfrac{\lambda}{2b}\left[\cos^2\phi+2\cos\phi\ln\left(\tan\dfrac{\phi}{2}\right)\right.\right.\notag\\
&\left.\left.-(1+\cos^2\phi)\ln\dfrac{\lambda\sin\phi}{b}\right]-\frac{q}{16 b^2}[9 \sin\phi+\sin3\phi+6 (\pi -2 \phi ) \cos\phi]+\mathcal{O}(\varepsilon^3)\right\}\label{PFDMuphi}
\end{align}
or
\begin{align}
\phi=&\arcsin b u+\frac{M \left(b^2 u^2-2\right)}{b \sqrt{1-b^2 u^2}}+\frac{M^2 \left(15 \arcsin(b u)-\frac{b u \left(3 b^4 u^4-20 b^2 u^2+15\right)}{\left(1-b^2 u^2\right)^{3/2}}\right)}{4 b^2}-\frac{q \left(b^3 u^3+3 \sqrt{1-b^2 u^2} \arcsin(b u)-3 b u\right)}{4 b^2 \sqrt{1-b^2 u^2}}\notag\\
&+\frac{\lambda  \left(b^2 u^2 (\ln(\lambda u)-1)+2 \sqrt{1-b^2 u^2} \ln \left(\frac{u}{\sqrt{1-b^2 u^2}+1}\right)-2 \ln (\lambda  u)+1\right)}{2 b \sqrt{1-b^2 u^2}}+\mathcal{O}\left(\varepsilon^3\right),\label{PFDMphivalue}
\end{align}
where we have dropped third or higher-order perturbation terms, and details on the derivation of (\ref{PFDMuphi}) is given in Appendix~\ref{A}.
Following the convention in~\cite{Ono:2019hkw}, we denote $\phi_S$ and $\phi_R$ as the value of $\phi$ at the source and the receiver, respectively, which satisfy $0\leq\phi_S<\frac{\pi}{2}$ and $\frac{\pi}{2}<\phi_R\leq\pi$.

Substituting (\ref{KRNdSPFDM}) and (\ref{elementRNdSPFDM}) into the first term of (\ref{finallyangle}), and taking only the first and second terms in (\ref{PFDMuphi}) and (\ref{PFDMphivalue}) for the integration limits\footnote{We only do the integral up to the second order $\mathcal{O}({\varepsilon^2})$. Note that the integrand has no zeroth-order term and begins with the first order. Therefore the second order terms in the integration limits only give rise to third- or higher-order terms, and hence are dropped. Furthermore, by carefully observing the original integrand one can check that its higher-order terms will not diverge after the integrals. Here for simplicity we abuse the jargon a bit, when we say the ``$n$th-order terms'' or $\mathcal{O}(\varepsilon^n)$, we actually have also included terms of $\mathcal{O}(\varepsilon^n\ln\varepsilon)$.}, the surface integral of the Gaussian curvature is performed as:
\begin{align}
\iint_{D_R+D_S}KdS=&\int_{\phi_S}^{\phi_R}\int_{r_0}^{r(\phi)}\left[-\frac{2M}{r^2}-\frac{3 M^2}{r^3}+\frac{\lambda  \left(2 \ln \left(\frac{r}{\lambda }\right)-3\right)}{2 r^2}+\frac{3q}{r^3}\right]drd\phi+\mathcal{O}\left(\varepsilon^3\right)\notag\\
=&\int_{\phi_S}^{\phi_R}\int_{u_0}^{u(\phi)}\left[2 M+3 M^2 u+\lambda  \left(\frac{3}{2}-\ln \left(\dfrac{1}{\lambda  u}\right)\right)-3 q u\right]dud\phi+\mathcal{O}\left(\varepsilon^3\right)\notag\\
=&\int_{\phi_S}^{\phi_R}\left[\dfrac{2M(\sin\phi-1)}{b}+\dfrac{M^2\cos^2\phi }{2b^2} +\dfrac{\lambda}{2b}  \left(\sin\phi\left[1-2\ln \left(\frac{b}{\lambda  \sin\phi}\right)\right]+2\ln \left(\frac{b}{\lambda }\right)-1\right)\right.\notag\\
&\left.+\frac{3q\cos^2\phi}{2b^2}\right]d\phi+\mathcal{O}\left(\varepsilon^3\right)\notag\\
=&\frac{2M}{b}\left[\sqrt{1-b^2 u_R^2}+\sqrt{1-b^2 u_S^2}\right]+\dfrac{M^2}{4b^2}\left[\frac{b u_R \left(15-7 b^2 u_R^2\right)}{\sqrt{1-b^2u_R^2}}+\frac{b u_S \left(15-7 b^2 u_S^2\right)}{\sqrt{1-b^2 u_S^2}}\right.\notag\\
&\left.+\arccos(bu_R)+\arccos(bu_S)\right]-\frac{\lambda}{2b}\left[\sqrt{1-b^2 u_R^2}+\sqrt{1-b^2 u_S^2}\right]\notag\\
&+\frac{\lambda}{b}\left\{\ln\left[\left(1+\sqrt{1-b^2 u_R^2}\right)\left(1+\sqrt{1-b^2 u_S^2}\right)\right]-\left(\sqrt{1-b^2 u_R^2}+\sqrt{1-b^2 u_S^2}\right)\ln\left(\dfrac{b}{\lambda}\right)\right.\notag\\
&\left.+\left(\sqrt{1-b^2u_R^2}-1\right)\ln(bu_R)+\left(\sqrt{1-b^2u_S^2}-1\right)\ln(bu_S)\right\}\notag\\
&-\frac{3q}{4b^2}\left[bu_R\sqrt{1-b^2 u_R^2}+bu_S\sqrt{1-b^2 u_S^2}\right]+\left(-\frac{2 M}{b}-\frac{\lambda}{2 b}  \left[1-2\ln \left(\frac{b}{\lambda }\right)\right]+\frac{3 q}{4 b^2}\right)\phi_{RS}+\mathcal{O}\left(\varepsilon^3\right).\label{DMGaussGcurvature}
\end{align}
Next, using (\ref{geodesiccurvature}), the integral of the geodesic curvature shown as the second term in (\ref{finallyangle}) is calculated as
\begin{align}
\int^{P_S}_{P_R}k_{g}(C_0)dl=&-\int^{\phi_R}_{\phi_S}k_{g}(C_0)\frac{dl}{d\phi}d\phi\notag\\
=&\left(-1+\frac{2 M}{b}-\frac{3 q}{4 b^2}+\frac{\lambda}{2 b}  \left[1-2\ln \left(\frac{b}{\lambda }\right)\right]\right)\phi_{RS}+\dfrac{7M^2}{2b^2}\left[\arccos(bu_R)+\arccos(bu_S)\right]\notag\\
&-\frac{3q}{4b^2}\left[\arccos(bu_R)+\arccos(bu_S)\right]
+\mathcal{O}\left(\varepsilon^3\right).\label{MDgeodesic}
\end{align}
Finally, putting together these results, we obtain the deflection angle
\begin{align}
\hat{\alpha}=&\frac{2M}{b}\left[\sqrt{1-b^2 u_R^2}+\sqrt{1-b^2 u_S^2}\right]+\dfrac{M^2}{4b^2}\left[\frac{b u_R \left(15-7 b^2u_R^2\right)}{\sqrt{1-b^2u_R^2}}+\frac{b u_S \left(15-7 b^2u_S^2\right)}{\sqrt{1-b^2 u_S^2}}+15 (\arccos(bu_R)+\arccos(bu_S))\right]\notag\\
&\notag\\
&-\frac{\lambda}{2b}\left[\sqrt{1-b^2 u_R^2}+\sqrt{1-b^2 u_S^2}\right]+\frac{\lambda}{b}\left\{\ln\left[\left(1+\sqrt{1-b^2 u_R^2}\right)\left(1+\sqrt{1-b^2 u_S^2}\right)\right]\right.\notag\\
&\left.-\left(\sqrt{1-b^2 u_R^2}+\sqrt{1-b^2 u_S^2}\right)\ln\left(\dfrac{b}{\lambda}\right)
+\left(\sqrt{1-b^2u_R^2}-1\right)\ln(bu_R)+\left(\sqrt{1-b^2u_S^2}-1\right)\ln(bu_S)\right\}\notag\\
&-\frac{3q}{4b^2}\left[bu_R\sqrt{1-b^2 u_R^2}+bu_S\sqrt{1-b^2 u_S^2}+\arccos(bu_R)+\arccos(bu_S)\right]+\mathcal{O}\left(\varepsilon^3\right).\label{DMangleGBT}
\end{align}
In Appendix~\ref{B}, we further demonstrate that our result of the deflection angle (\ref{DMangleGBT}) is consistent with the one from doing direct integral along the geodesic.

As an additional remark, when the source and the receiver are so far away from the black hole such that $bu_S$ and $bu_R$ are of the order $\mathcal{O}(\varepsilon)$ or higher, in (\ref{DMangleGBT}) one can send $u_R\rightarrow 0$ and $u_S\rightarrow 0$ without affecting the leading order terms, which gives\footnote{Note that by sending $u_{R,S}\rightarrow0$ this result agrees with the calculation in Ref.~\cite{Atamurotov:2021hck}, except that the authors there did not involve the order $\mathcal{O}(M^2/b^2)$ into the calculation.}
\begin{align}
\hat{\alpha}= \frac{4M}{b}+\dfrac{15\pi M^2}{4b^2}-\frac{3\pi q}{4b^2}-\frac{\lambda}{b}\left(1+2\ln\frac{b}{\lambda}-2\ln2\right)+\mathcal{O}\left(\varepsilon^3\right).\label{approximationangle}
\end{align}

\section{The size of the Einstein ring}
\label{section4}
In this section, we derive the analytical expression of the angular radius of the Einstein's ring in the charged black hole immersed in PFDM. We only consider the special situation that the source, lens and receiver are aligned along the same axis, and we assume that the source and receiver are sufficiently far so that (\ref{approximationangle}) holds.

\begin{figure}[h]
\centering
\includegraphics[width=3.2in]{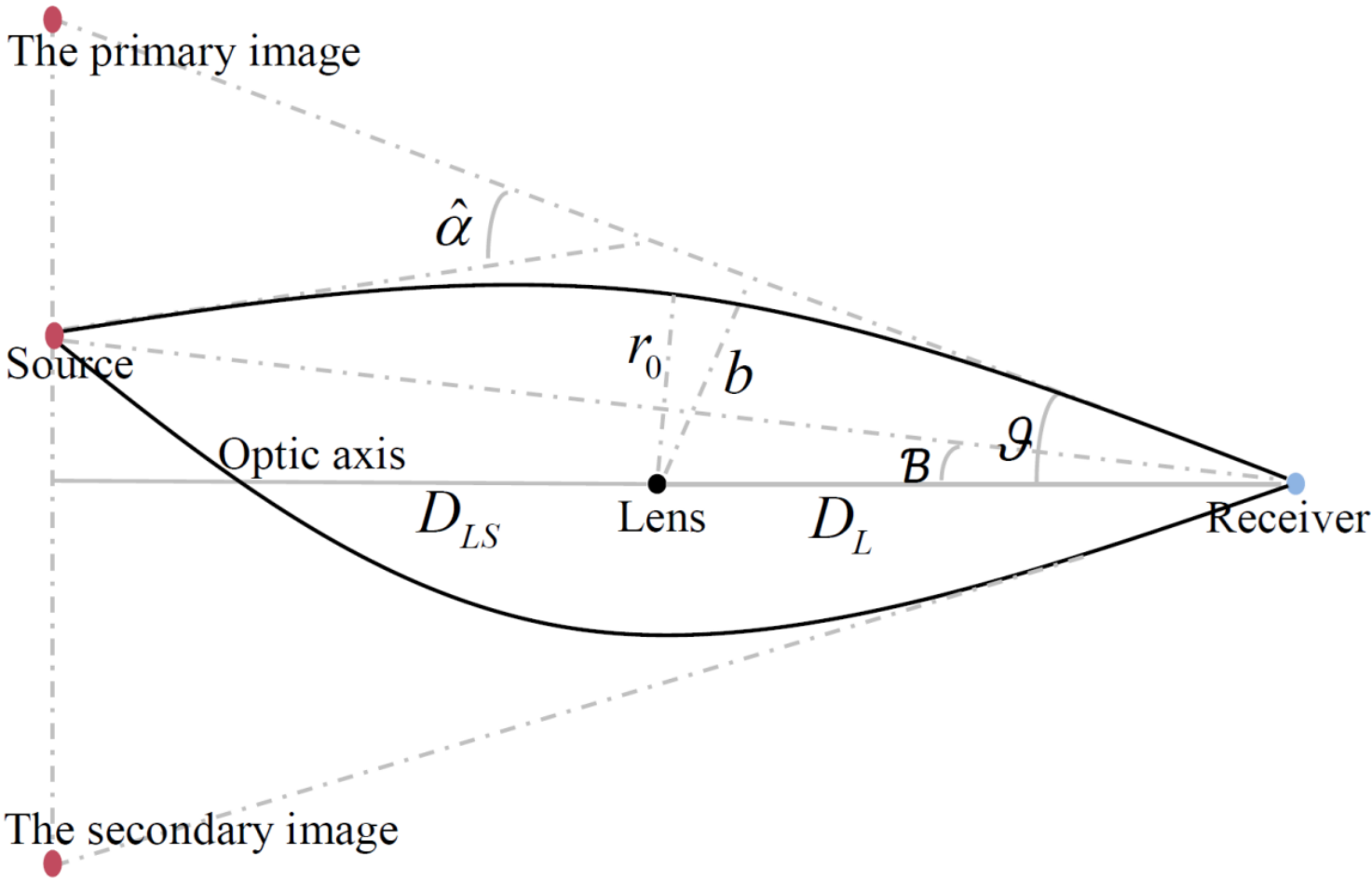}
\caption{The schematic diagram of light bending and GL.}
\label{lenspicture}
\end{figure}

Bozza derived in \cite{Bozza:2008ev} the lensing equation that relates the angular position of the unlensed source with that of its image (denoted by $\mathcal{B}$ and $\vartheta$ as indicated in Fig.~\ref{lenspicture}):
\begin{align}
D_S\tan\mathcal{B}=\dfrac{D_L\sin\vartheta-D_{LS}\sin(\hat{\alpha}-\vartheta)}{\cos(\hat{\alpha}-\vartheta)},\label{Bozzalensequation}
\end{align}
where $D_L$ denotes the angular diameter distance from the receiver to the lens plane; $D_{LS}$ denotes the angular diameter distance from the lens plane to the source plane; $D_S=D_{LS}+D_L$ and furthermore $\sin(\vartheta)=b/D_L$.

In the situation that the source, lens and receiver are aligned we set $\mathcal{B}=0$, and thus in the weak deflection approximation\footnote{One can see from (\ref{approximationangle}) and (\ref{Bozzalensequation}) with $\mathcal{B}=0$ that $\hat \alpha$ and $\vartheta$ are of the order $\mathcal{O}(\varepsilon)$.}
we can solve (\ref{Bozzalensequation}) for $\vartheta$ as:
\begin{align}
\vartheta_E= \frac{D_{LS}}{D_S}\hat{\alpha}+\mathcal{O}\left(\varepsilon^3\right),\label{Einsteinring}
\end{align}
which gives the Einstein's angle denoted by $\vartheta_E$.
Substituting (\ref{approximationangle}) and
\begin{align}
\vartheta_E = \arcsin\left(\dfrac{b}{D_L}\right) = \dfrac{b}{D_L} + \mathcal{O}\left(\varepsilon^3\right)
\end{align}
into (\ref{Einsteinring}), we obtain
\begin{align}
\dfrac{b}{D_L}=\dfrac{D_{LS}}{D_S}\left[\dfrac{4M}{b}+\dfrac{15\pi M^2}{4 b^2}-\dfrac{3\pi q}{4b^2}-\dfrac{\lambda}{b}\left(1+2\ln\left(\dfrac{b}{D_L}\right)+2\ln\left(\dfrac{D_L}{\lambda}\right)-2\ln2\right)\right]+ \mathcal{O}\left(\varepsilon^3\right).\label{new-expression}
\end{align}
or equivalently
\begin{align}
\vartheta_E =\dfrac{D_{LS}}{D_S{\vartheta_E}}\left[\dfrac{4M}{D_L}-\dfrac{\lambda}{D_L}\left(1+\ln\vartheta_E^2+2\ln\dfrac{D_L}{\lambda}-2\ln2\right)\right]+\dfrac{D_{LS}}{D_S\vartheta_E^2}\left[\dfrac{15\pi M^2}{4D_L^2}-\dfrac{3\pi q}{4 D_L^2}\right]+ \mathcal{O}\left(\varepsilon^3\right).\label{new-vartheta2}
\end{align}

In the following we would like to solve (\ref{new-vartheta2}) for $\vartheta_E$. The existence of the logarithmic term $\ln \vartheta_E^2$ on right hand side hinders us from obtaining an analytical solution, but note that from the first-order terms in (\ref{new-vartheta2}) we can derive
\begin{align}
 \vartheta_E^2=\dfrac{D_{LS}}{D_S}\dfrac{4M}{D_L} +\mathcal{O}\left(\varepsilon^3\right),\label{b2DLexpression}
\end{align}
which further gives
\begin{align}
\ln\vartheta_E^2=\ln\left(\dfrac{D_{LS}}{D_S}\dfrac{4M}{D_L} \right)+\mathcal{O}(\varepsilon),\label{lntheta}
\end{align}
Thus we can use (\ref{lntheta}) to eliminate the problematic logarithm\footnote{In Ref.~\cite{Atamurotov:2021hck}, this type of problematic logarithm was circumvented in a different approach: $\ln(b/\lambda)$ was replaced with $\ln(10^n)$, with $n$ being a new parameter introduced by hand. Note that in our paper here we do not rely on any new artificial parameter -- instead we do more careful approximations by making sure that everything we change or drop is beyond the second order.} in (\ref{new-vartheta2}), and then equation can be further rewritten as:
\begin{align}
\vartheta_E^3=\dfrac{D_{LS}}{D_S}(\delta_2+\eta_3)\vartheta_E+\dfrac{D_{LS}}{D_S{}}\xi_4+ \mathcal{O}\left(\varepsilon^5\right),\label{simplifyvartheta}
\end{align}
in which
\begin{align}
\delta_2=\dfrac{4M}{D_L},~~\eta_3=-\dfrac{\lambda}{D_L}\left(1+\ln\left(\dfrac{4MD_{LS}}{D_L D_S}\right)+2\ln\dfrac{D_L}{\lambda}-2\ln2\right),~~\xi_4=\dfrac{15\pi M^2}{4D_L^2}-\dfrac{3\pi q}{4 D_L^2},\label{threecoefficients}
\end{align}
where each number in the subscript stands for the order, i.e.\ $\delta_2\sim\mathcal{O}(\varepsilon^2)$, $\eta_3\sim\mathcal{O}(\varepsilon^3)$ and $\xi_4\sim\mathcal{O}(\varepsilon^4)$.
We write $\vartheta_E$ in different orders as:
\begin{align}
\vartheta_E=\vartheta_{E1}+\vartheta_{E2}+\mathcal{O}\left(\varepsilon^3\right),\label{expandvartheta}
\end{align}
then we can solve (\ref{simplifyvartheta}) order by order, i.e.\ solve
\begin{align}
\vartheta_{E1}^3=&\dfrac{D_{LS}}{D_S}\delta_2\vartheta_{E1},\label{equation1}\\
3\vartheta_{E1}^2\vartheta_{E2}=&\dfrac{D_{LS}}{D_S}(\delta_2\vartheta_{E2}+\eta_3\vartheta_{E1}+\xi_4),\label{eqyation2}
\end{align}
which gives
\begin{align}
\vartheta_{E1}=\sqrt{\dfrac{D_{LS}}{D_S}\delta_2},~~\vartheta_{E2}=\dfrac{1}{2}\left(\sqrt{\dfrac{D_{LS}}{D_S}}\dfrac{\eta_3}{\sqrt{\delta_2}}+\dfrac{\xi_4}{\delta_2}\right).\label{twocoefficients}
\end{align}
Finally, substituting (\ref{threecoefficients}) into (\ref{twocoefficients}) and then into (\ref{expandvartheta}), we obtain an analytical expression of the angular radius of the Einstein ring
\begin{align}
\vartheta_E=\sqrt{\dfrac{4D_{LS}M}{D_SD_L}}+\dfrac{1}{4}\left[\dfrac{3\pi(5M^2-q)}{8D_L M}-\dfrac{\lambda \sqrt{D_{LS}}}{\sqrt{D_SD_L M}}\left(1+\ln\dfrac{D_{LS}D_L M}{D_S\lambda^2}\right)\right]+ \mathcal{O}\left(\varepsilon^3\right).\label{vartheta-value}
\end{align}

\section{Conclusion and discussion}
\label{section5}
In this paper, using the method based on the Gauss-Bonnet theorem, we have analytically calculated the gravitational lensing effect of a charged spherically symmetric black hole immersed in PFDM. Our calculation have been done in the weak field approximation, i.e.\ the ratio between the mass parameter $M$ and the impact parameter $b$ is small (which in our convention is a first-order small quantity), and furthermore we have assumed that the dark matter parameter $\lambda$ and the charge parameter $q$ only give second-order small contributions to the result. We have calculated the light deflection angle carefully up to second-order approximation, with the source and receiver placed at finite distances from the black hole. In the situation that the impact parameter is much smaller than the distance of the source (receiver), i.e.\ their ratio is no more than a first-order small quantity, it is a good approximation to treat the source (receiver) to be infinitely far. In this situation we have further derived the analytical expression of the Einstein's angular radius for the aligned source, lens and receiver, where again approximations (especially for the logarithmic term) have been carefully treated up to the second order.

The question that still remains is to what kind of celestial bodies our calculation may be applicable. For example, if we want to fit our analytical formulas correctly with real dark-matter lensing effect in observations, the celestial bodies in question must satisfy $\mathcal{O}(\lambda/b)\sim\mathcal{O}(M^2/b^2)$. That means in principle the calculation in this paper should only be applied to those astronomical systems where the presence of dark matter is much less abundant than conventional things that contribute to the mass parameter. Therefore, we think observations of lensing effects of regular stars and black holes should be potential playgrounds to apply the result of this paper, but observations of galaxies or their clusters should be much less relevant due to the dark-matter dominance there (except perhaps a few special cases like \cite{vanDokkum:2018vup}). Note however that there may be some subtleties in the physical interpretation of the parameters $M$ and $\lambda$ for a galaxy (cluster). On the one hand, it seems most reasonable to consider $\lambda$ as the only parameter responsible for dark-matter contribution and $M$ as anything else like luminous matter and black holes; on the other hand, since the nature of dark matter has never been well-understood, perhaps it is possible that $M$ also includes contribution from the dark matter within the galaxy (cluster) and $\lambda$ acts as an additional correction contributed by the dark matter around it. In the former case, it is likely that $\lambda/b$ is way much larger than a second-order small quantity, and in the latter case, usually the precision of current observations is not sufficient to tell the order of magnitude of $\lambda$, which we have used a concrete example in Appendix \ref{C} to illustrate. In principle, we may generalize the calculation in this paper to the ``thick'' dark matter case. In this case we must consider higher order terms of $\lambda$, but then the integral cannot
be easily done analytically due to the logarithmic terms both in the integrand and in the integration limits. This is the reason why we have restricted our discussion to the ``thin'' case only, but the ``thick'' case is for sure interesting for future investigations. Recently, a paper \cite{Qiao:2022nic} appears to be working on the ``thick'' dark matter case, but unfortunately their result seems to be incorrect, since it does not reproduce the coefficient of the $M^2$-term in Eq.(24) of \cite{Keeton:2005jd}.
\footnote{We thank the anonymous referee for reminding us of \cite{Qiao:2022nic}.
We would like to point out that the inconsistency between \cite{Qiao:2022nic} and \cite{Keeton:2005jd} comes from Eq.(49) and (56) in the second arXiv version of \cite{Qiao:2022nic}. There the integration limits are only approximated up to the zeroth-order, while first-order terms in the limits are ignored, which actually do have contributions to the results at the second order and should be taken into account.
In addition, it seems that the difficulty in calculating the ``thick'' case has been underestimated by the authors of \cite{Qiao:2022nic}.}

\section{Acknowledgements}
We are grateful to Profs. Hongsheng Zhang and Drs. Shi-Bei Kong, Tao-Tao Sui, Arshad
Ali and Yu-Ting Zhou for interesting and stimulating discussions. This work is supported by National Natural Science Foundation of China (NSFC) under grant Nos. 12175105, 11575083, 11565017, 12147175, Top-notch Academic Programs Project of Jiangsu Higher Education Institutions (TAPP).

\appendix
\section{Details for the derivation of (\ref{PFDMuphi})}\label{A}
In this appendix, we give the detailed derivation of (\ref{PFDMuphi}). For clarity we define $\bar{M}$, $\bar{\lambda}$, $\bar{q}$ and $\bar{u}$ as the following:
\begin{align}
\dfrac{M}{b}\equiv \varepsilon \bar{M},~ \dfrac{\lambda}{b}\equiv \varepsilon^2 \bar{\lambda},~\dfrac{q}{b^2}\equiv \varepsilon^2 \bar{q},~bu\equiv\bar{u}.\label{parametersetting}
\end{align}
Note that these barred parameters are all dimensionless quantities of $\mathcal{O}(\varepsilon^0)$. Then we rewrite (\ref{RNdSPFDMtrajectory}) as
\begin{align}
\left(\dfrac{d\bar{u}}{d\phi}\right)^2+\bar{u}^2=1+2\bar{u}^3\varepsilon \bar{M}+\bar{u}^3\ln(\varepsilon^2\bar{\lambda}\bar{u})\varepsilon^2\bar{\lambda}
-\bar{u}^4\varepsilon^2\bar{q}.\label{newmotioneq}
\end{align}
Taking the derivative of both sides of (\ref{newmotioneq}) with respect to $\phi$, we can obtain
\begin{align}
\dfrac{d^2\bar{u}}{d\phi^2}+\bar{u}=3\bar{u}^2\varepsilon\bar{M}+\dfrac{\bar{u}^2}{2}[1+3\ln(\varepsilon^2\bar{\lambda}\bar{u})]\varepsilon^2\bar{\lambda}
-2\bar{u}^3\varepsilon^2\bar{q}.\label{newsecondderiveeq}
\end{align}
We expand $\bar{u}$ in a series:
\begin{align}
\bar{u}=\bar{u}_0+\varepsilon \bar{u}_1+\varepsilon^2 \bar{u}_2+\mathcal{O}(\varepsilon^3),\label{uexpand}
\end{align}
and substitute (\ref{uexpand}) into (\ref{newmotioneq}) and into (\ref{newsecondderiveeq}), respectively, which leads to the following two set of equations:
\begin{align}
\left\{
    \begin{array}{lc}
    \varepsilon^0:~\left(\dfrac{d\bar{u}_0}{d\phi}\right)^2+\bar{u}_0^2=1,\\
    \varepsilon^1:~\dfrac{d\bar{u}_1}{d\phi}\dfrac{d\bar{u}_0}{d\phi}+\bar{u}_1\bar{u}_0=\bar{u}_0^3\bar{M},\\
    \varepsilon^2:~2\dfrac{d\bar{u}_2}{d\phi}\dfrac{d\bar{u}_0}{d\phi}+\left(\dfrac{d\bar{u}_1}{d\phi}\right)^2
    +2\bar{u}_2\bar{u}_0+\bar{u}_1^2=\bar{\lambda}\bar{u}_0^3\ln(\varepsilon^2\bar{\lambda}\bar{u}_0)
    +6\bar{u}_0^2\bar{u}_1\bar{M}-\bar{u}_0^4\bar{q},\\
    \cdot\cdot\cdot
    \end{array}
\right.\label{threefirstdermotion}
\end{align}
and
\begin{align}
\left\{
    \begin{array}{lc}
   \varepsilon^0:~ \dfrac{d^2\bar{u}_0}{d\phi^2}+\bar{u}_0=0,\\
     \varepsilon^1:~\dfrac{d^2\bar{u}_1}{d\phi^2}+\bar{u}_1=3\bar{u}_0^2\bar{M},\\
    \varepsilon^2:~\dfrac{d^2\bar{u}_2}{d\phi^2}+\bar{u}_2=6b\bar{u}_0\bar{u}_1\bar{M}
    +\dfrac{\bar{u}_0^2}{2}[1+3\ln(\varepsilon^2\bar{\lambda}\bar{u}_0)]\bar{\lambda}-2\bar{u}_0^3\bar{q},\\
    \cdot\cdot\cdot
    \end{array}\label{threeseconddermotion}
\right.
\end{align}
We choose $\dfrac{d\bar{u}}{d\phi}\big|_{\phi=\pi/2}=0$ as the boundary condition, and then at $\phi=\pi/2$ we can algebraically solve (\ref{threefirstdermotion}) as
\begin{align}
\bar{u}_0\left(\dfrac{\pi}{2}\right)=1,~\bar{u}_1\left(\dfrac{\pi}{2}\right)=\bar{M},~\bar{u}_2\left(\dfrac{\pi}{2}\right)=\dfrac{5\bar{M}-\bar{q}+\bar{\lambda}\ln (\varepsilon^2\bar{\lambda})}{2}.\label{initialconditions}
\end{align}
Using (\ref{initialconditions}) as the boundary condition, we can solve (\ref{threeseconddermotion}) order by order to obtain
\begin{align}
\bar{u}_0&=\sin (\phi ),~\bar{u}_1=\bar{M}(1+\cos^2\phi),\label{u0uisolution}\\
\bar{u}_2&=-\dfrac{\bar{\lambda}}{2}\left[\cos^2\phi+2\cos\phi\ln\left(\tan\dfrac{\phi}{2}\right)-\ln(\varepsilon^2\bar{\lambda}\sin\phi)(1+\cos^2\phi)
\right]-\frac{\bar{q} [9 \sin (\phi )+\sin (3 \phi )+6 (\pi -2 \phi ) \cos (\phi )]}{16 }\notag\\
&+\frac{\bar{M}^2 [37 \sin (\phi )-3 \sin (3 \phi )+30 (\pi -2 \phi ) \cos (\phi )]}{16 },\label{u2solution}
\end{align}
and hence
\begin{align}
\bar{u}=&\sin (\phi )+\varepsilon\bar{M}(1+\cos^2\phi)-\dfrac{\varepsilon^2\bar{\lambda}}{2}\left[\cos^2\phi+2\cos\phi\ln\left(\tan\dfrac{\phi}{2}\right)-\ln(\varepsilon^2\bar{\lambda}\sin\phi)(1+\cos^2\phi)
\right]\notag\\
&-\frac{\varepsilon^2\bar{q} [9 \sin (\phi )+\sin (3 \phi )+6 (\pi -2 \phi ) \cos (\phi )]}{16}+\frac{\varepsilon^2\bar{M}^2 [37 \sin (\phi )-3 \sin (3 \phi )+30 (\pi -2 \phi ) \cos (\phi )]}{16}+\mathcal{O}(\varepsilon^3),\label{pertursolu}
\end{align}
which via (\ref{parametersetting}) further leads to (\ref{PFDMuphi}).

\section{Calculation of deflection angle with the integral method}\label{B}
In this appendix, as a verification of the former results (\ref{DMangleGBT}), we derive the weak deflection angle of light by using the integral method~\cite{Ishihara:2016vdc} .

First, from (\ref{sinPsi}), we easily obtain
\begin{align}
\Psi_R-\Psi_S=&\arcsin(bu_R)+\arcsin(bu_S)-\pi-\frac{M}{b}\left(\frac{b^2 u_R^2}{\sqrt{1-b^2 u_R^2}}+\frac{b^2 u_S^2}{\sqrt{1-b^2 u_S^2}}\right)\notag\\
&+\dfrac{M^2}{2b^2}\left(\frac{b^3 u_R^3 \left(2 b^2 u_R^2-1\right)}{\left(1-b^2 u_R^2\right)^{3/2}}+\frac{b^3 u_S^3 \left(2 b^2 u_S^2-1\right)}{\left(1-b^2u_S^2\right)^{3/2}}\right)\notag\\
&+\frac{q}{b^2}\left(\frac{b^3u_R^3}{2 \sqrt{1-b^2u_R^2}}+\frac{b^3u_S^3}{2 \sqrt{1-b^2u_S^2}}\right)-\frac{\lambda}{b}\left(\frac{b^2u_R^2 \ln\left(\lambda u_R\right)}{2 \sqrt{1-b^2u_R^2}}+\frac{b^2u_S^2 \ln\left(\lambda u_S\right)}{2 \sqrt{1-b^2u_S^2}}\right)+\mathcal{O}\left(\varepsilon^3\right).\label{DMPsiRS}
\end{align}
Next, by making use of (\ref{RNdSPFDMtrajectory}), we have
\begin{align}
\phi_{RS}=&\int_{u_S}^{u_0}\dfrac{du}{\sqrt{F(u)}}+\int_{u_0}^{u_R}\dfrac{du}{\sqrt{F(u)}}\notag\\
=&\pi-\arcsin(bu_R)-\arcsin(bu_S)+\frac{M}{b}\left(\frac{2-b^2u_R^2}{\sqrt{1-b^2u_R^2}}+\frac{2-b^2u_S^2}{\sqrt{1-b^2u_S^2}}\right)\notag\\
&+\dfrac{M^2}{4b^2}\left(\frac{bu_R \left(3 b^4u_R^4-20 b^2u_R^2+15\right)}{\left(1-b^2u_R^2\right)^{3/2}}+\frac{b u_S \left(3 b^4 u_S^4-20 b^2 u_S^2+15\right)}{\left(1-b^2 u_S^2\right)^{3/2}}+15(\arccos(bu_R)+\arccos(bu_S))\right)\notag\\
-&\frac{q}{4b^2}\left(3[\arccos(bu_R)+\arccos(bu_S)]+\frac{bu_R\left(3-b^2u_R^2\right)}{ \sqrt{1-b^2u_R^2}}+\frac{bu_S\left(3-b^2u_S^2\right)}{\sqrt{1-b^2u_S^2}}\right)\notag\\
&-\frac{\lambda}{2b}\left[\left(\sqrt{1-b^2u_R^2}+\sqrt{1-b^2u_R^2}\right)\right]+\frac{\lambda}{b}\left\{\ln\left[ \left(\sqrt{1-b^2u_R^2}+1\right)\left(\sqrt{1-b^2u_S^2}+1\right)\right]\right.\notag\\
&\left.+\frac{1}{2}\left(\frac{(2-b^2u_R^2)\ln(\lambda u_R)}{\sqrt{1-b^2u_R^2}}+\frac{(2-b^2u_S^2)\ln(\lambda u_S)}{\sqrt{1-b^2u_S^2}}\right)-\ln(bu_R)-\ln(bu_S)\right\}+\mathcal{O}\left(\varepsilon^3\right).\label{DMpsiRS}
\end{align}
Substituting (\ref{DMPsiRS}) and (\ref{DMpsiRS}) into (\ref{definitionangle}) we obtain the expression of the deflection angle
\begin{align}
\hat{\alpha}=&\frac{2M}{b}\left(\sqrt{1-b^2 u_R^2}+\sqrt{1-b^2 u_S^2}\right)-\frac{3q}{4b^2}\left(bu_R\sqrt{1-b^2 u_R^2}+bu_S\sqrt{1-b^2 u_S^2}+\arccos(bu_R)+\arccos(bu_S)\right)\notag\\
&+\dfrac{M^2}{4b^2}\left(\frac{b u_R \left(15-7 b^2u_R^2\right)}{\sqrt{1-b^2u_R^2}}+\frac{b u_S \left(15-7 b^2u_S^2\right)}{\sqrt{1-b^2 u_S^2}}+15 \left(\arccos(bu_R)+\arccos(bu_S)\right)\right)\notag\\
&-\frac{\lambda}{2b}\left(\sqrt{1-b^2 u_R^2}+\sqrt{1-b^2 u_S^2}\right)+\frac{\lambda}{b}\left\{\ln\left[\left(1+\sqrt{1-b^2 u_R^2}\right)\left(1+\sqrt{1-b^2 u_S^2}\right)\right]\right.\notag\\
&\left.-\left(\sqrt{1-b^2 u_R^2}+\sqrt{1-b^2 u_S^2}\right)\ln\left(\dfrac{b}{\lambda}\right)
+\left(\sqrt{1-b^2u_R^2}-1\right)\ln(bu_R)+\left(\sqrt{1-b^2u_S^2}-1\right)\ln(bu_S)\right\}\notag\\
&+\mathcal{O}\left(\varepsilon^3\right),\label{DMangleIshihara}
\end{align}
which is exactly the same as the result (\ref{DMangleGBT}) that is derived by using the Gauss-Bonnet theorem.

\section{Estimating the magnitude of $\lambda$ from the observational data of the galaxy ESO325-G004}
\label{C}
In this section, we try to use the data of galaxy ESO325-G004 to estimate the order of magnitude of $\lambda$ to illustrate our discussion at the end of Section \ref{section5}. All the data below are from \cite{Smith:2013ena} and its references, and for the purpose of our discussion we only do very rough estimations without error analysis.

The observed value of the Einstein ring around the center of the galaxy ESO325-G004 is
\begin{equation}
    \vartheta_E^{obs}=2.85''\label{vartheta-ESO} .
\end{equation}
The lens redshift is $z_l=0.035$; and the source (the background galaxy) redshift is $z_s=1.141$. Hubble has found that the relation between the spectrum redshift and distance of galaxy i.e. $z=H_0 d$, where $H_0=70.4\, {\rm km}\, {\rm s}^{-1}\,({\rm Mpc})^{-1}$ is the Hubble constant and $d$ denotes the proper distance, which is equal to the scale factor $a$ multiply the comoving distance $D$, i.e. $d=aD$ with $a=1/(1+z)$. Then we have
\begin{align}
D_S=\dfrac{c z_s(1+z_s)}{H_0}=2.864\times10^4\,{\rm Mpc}, \quad D_L=\dfrac{c z_l(1+z_l)}{H_0}=1.543\times10^2\,{\rm Mpc} \ .\label{ESOdata}
\end{align}

According to \cite{Smith:2013ena}, the mass parameter $M$ for the galaxy ESO325-G004 is about $10^{14} {\rm m}$, and thus $M/b\sim10^{-5}$, where $b=D_L\sin\vartheta_E^{obs}\sim10^{19}{\rm m}$. Therefore, the formulas in this paper is applicable only when $\lambda/b \sim \mathcal{O}(M^2/b^2) \sim 10^{-10}$, i.e.\  $\lambda \sim 10^{9} \rm{m}$.

We assume that the galaxy is neutral in charge and thus (\ref{vartheta-value}) is simplified as
\begin{align}
\vartheta_E=\sqrt{\dfrac{4D_{LS}M}{D_SD_L}}+\dfrac{1}{2}\left[\dfrac{3\pi 5M}{16D_L}-\dfrac{\lambda D_{LS}}{2\sqrt{D_{LS}D_SD_L M}}\left(1+\ln\left(\dfrac{D_{LS}D_L M}{D_S\lambda^2}\right)\right)\right]+ \mathcal{O}\left(\varepsilon^3\right),\label{noq-vartheta-value}
\end{align}
which by substituting (\ref{vartheta-ESO}) and (\ref{ESOdata}) gives an equation that relates $M$ and $\lambda$.
Then by inputting some value of $\lambda$, $M$ can be obtained numerically, and vice versa. For example, if we let $\lambda$ run from 0 to $10^{11}{\rm m}$, $M$ varies from $1.546\times10^{11}M_{\odot}$ to $1.554\times10^{11}M_{\odot}$, where $M_{\odot}=1.989\times 10^{30} {\rm kg}$ is the solar mass.

As discussed at the end of Section \ref{section5}, there may be different ways to interpret $M$. Here if we set $M$ to be the mass of the luminescent matter $M_L=1.28^{+0.10}_{-0.17} \times 10^{11} M_{\odot}$ \cite{Smith:2013ena}, the corresponding $\lambda$ has the order of magnitude $10^{19} \rm{m}$, which means $\lambda/b$ in this case has to be much larger than a second-order small value, and thus the result in this paper is not applicable. On the other hand, if we set $M$ to be the total mass of both the luminescent and dark matter $M_{L+D}=1.50\pm 0.06 \times 10^{11} M_{\odot}$ \cite{Smith:2013ena}, then $\lambda$ can run from 0 to $10^{11}{\rm m}$ without exceeding the allowed range of observational error, and that is to say, there may be a chance that $\lambda/b$ is of the second order, but the current data is not sufficiently precise to conclude this.

\end{document}